\documentstyle[epsfig,12pt]{article}

\newcommand{\beq}{\begin{equation}}
\newcommand{\eeq}{\end{equation}}

\begin{document}
\def\lag{\langle}
\def\rag{\rangle}

 \title{Controlling a Non-Polynomial \\
Reduced Finite Temperature Action\\
in the U(1) Higgs Model\\
}

\author{{A. Jakov\'ac$^{1}$, A. Patk\'os$^{2}$ and P. Petreczky$^{3}$}\\
{Department of Atomic Physics}\\
{E\"otv\"os University, Budapest, Hungary}\\
}
\vfill
\footnotetext[1]{{\em e-mail address: jako@hercules.elte.hu}}
\footnotetext[2]{{\em e-mail address: patkos@ludens.elte.hu}}
\footnotetext[2]{{\em e-mail address: ppetrecz@dragon.klte.hu}}
\maketitle
\begin{abstract}
An effective theory is constructed for the scalar electrodynamics
via 2-loop integration over all non-static 
fields and the screened electric component of the vector-potential.
 Non-polynomial terms of the
action are preserved and included into the 2-loop calculation
of the effective potential of the reduced theory. Also the inclusion of
some non-local terms is shown to be important. The effect
of non-polynomial operators on the symmetry restoring phase transition
is quantitatively compared to results
from a local, superrenormalisable approximate effective theory.
\end{abstract}
\newpage
1. The thermodynamics of high temperature phases and of
 finite temperature phase transitions of gauge theories is actively studied
with help of reduced three-dimensional effective theories 
\cite{kajantie1,braaten1,kripfganz}. In the construction
of the actions of these effective theories substantial progress has been made
recently, based on matching the effective theories to the original full finite
temperature theory \cite{farakos1,kajantie2}. In this strategy one
computes a set of observables in both theories and fixes the relation
of the couplings by requiring their agreement.
 In the high temperature perturbative regime this approach met considerable 
success \cite{braaten2}. 

Non-perturbative investigations of full finite temperature theories
\cite{csikor,montvay,karsch} represent important reference points to every
effective model proposition.
 
Our aim is to investigate a related but conceptually different procedure
 to arrive at effective models: the partial integration over a set of field 
variables. This "identity"-transformation should allow 
to check the accuracy of some
physically very appealing candidates for the effective theories of finite
temperature phase transitions.
  
Integration over non-static fields leads to three-dimensional 
representation of finite temperature field theories. Since the 
 removal of four-dimensional singularities can be fully implemented
in this step of partial integration, and in the full theory no intrinsic 
three-dimensional divergences are present, the emerging theory is 
{\it finite}.
This means that any three-dimensional cut-off dependence appearing 
in the calculation
of correlation functions from the reduced theory, will be cancelled exactly
by the cut-off dependence of its couplings, induced in the step of partial 
integration of non-static field variables (PI-step). 
                           
In view of the above feature, three-dimensional renormalisability 
of the reduced model
is not required, the presence of higher dimensional
or even non-polynomial terms in the reduced action is equally well
admissible. The process one follows in the perturbative approach
to the reduced theory consists of the following steps:

i) Separation of all couplings of the effective model
into a finite and a cut-off dependent part,
e.g.
\begin{equation}
M_{3D}^2(\Lambda ,T, m_{4D}^2,...)=m_{3D}^2(T,m_{4D}^2,...)+c_1\Lambda T+
c_2\Lambda T\log {\Lambda\over \mu}+c_3 T^2\log {\Lambda\over \mu},
\end{equation}
etc.

ii) Perturbative calculation  of physical quantities with the finite 
parts of the couplings, and using the cut-off dependent part of 
the action as counterterm. 

iii) Separation of the cut-off independent part
of the result from the divergent pieces
(for a unique separation one has to choose in the course of solving the 
reduced theory the same scale $\mu$ as introduced above).
 Since the cancellation of the divergent pieces 
is exact, the finite part provides the  physical answer.

It is interesting to note that in case of the effective potential 
 the three-dimensional 1-loop "counterterm" contribution has no
finite part, therefore the physical answer (at least up
to 2-loop) is simply the finite part of the
perturbative contribution calculated with the finite parts of the couplings.

In the present note we shall work out explicitly for the U(1) Higgs model
one particularly  appealing version of the reduced theory. It
arises when also the dynamically screened electric vector-potential
component 
is included into the PI-step. The 2-loop accurate integration will follow
the procedure of the gradient expansion. When calculating the local 
(potential) term of the reduced action we shall find and retain an
 ${\cal O}(e^3)$non-polynomial contribution. We are able to show that its 
cut-off dependent part is exactly cancelled when the effective theory
is solved. 

In the next step the second derivative (kinetic)
part of the reduced action is determined by calculating the T-dependent
wave function renormalisation of the static scalar and magnetic vector
fields, due to the integrated out fields. 
It turns out that truncating the effect of the
screened electric vector component  at this stage of the gradient expansion  
leads to a not fully satisfactory 2-loop effective potential. 
 The formal expression  of the 
2-loop effective potential calculated directly from the 
four-dimensional full theory \cite{hebecker1,hebecker2} 
can be reproduced only if
a bilocal term of non-locality range $eT$ is taken to represent the effect 
of the $A_0$-integration.  

A quantitative comparison will be made between the above complete 2-loop
treatment and the approach in which the non-polynomial
part of the potential is expanded up to quartic power in the Higgs-field
(super-renormalisable
approximation). Calculating some data of the first order phase 
transition restoring the U(1) symmetry with both approaches, 
one can assess within the perturbation theory 
the impact of the higher dimensional operators.
\vskip 1truecm
2. The model is defined with the following Lagrangian density:
\begin{eqnarray}
&
L=L_{ren}+L_{ct},   \nonumber\\
&
L_{ren}={1\over 4}F_{mn}F_{mn}+{1\over 2}(D_m\phi )^{*}(D_m\phi )+
{1\over 2}m^2\phi^*\phi\nonumber\\
&
+{1\over 24}\lambda(\phi^*\phi )^2+{1\over 2}m_D^2
A_0^2({\bf x}),\nonumber\\
&
L_{ct}={1\over 2}\delta Z_AF_{mn}F_{mn}+{1\over 2}\delta Z_{\phi^2}
(D_m\phi )^*(D_m\phi ) +{1\over 2}(\delta m^2+\delta Z_{\phi^2}m^2)
\phi^*\phi \nonumber\\
&
+ {1\over 24}(\delta\lambda +2\lambda\delta 
Z_{\phi^2})(\phi^*\phi )^2-{1\over 2}m_D^2A_0^2({\bf x}),
\label{lagrange}
\end{eqnarray}
with
\begin{equation}
F_{mn}=\partial_mA_n-\partial_nA_m,~~~D_m\phi =(\partial_m+ieA_m)\phi,~~~
m,n=1,..,4
\end{equation}
One notes in (\ref{lagrange}) 
the mass term for the static $A_0({\bf x})$ field reflecting its 
Debye-screened
nature. All couplings and fields appearing in (\ref{lagrange}) are renormalised 
quantities.

The contribution of the non-static fluctuations to the local potential of the 
static $\phi$-fields can be evaluated up to  2-loops using exactly those 
diagrams which appear in the direct evaluation of the effective potential
\cite{arnosa}. We also use Landau-gauge.
The explicit expression is formally the same in 
terms of two fundamental  sum-integrals. The essential difference is the 
absence of the $n=0$ mode from the sum. As a consequence one can expand 
these integrals with respect to the mass(es):
\begin{eqnarray}
&
\int_{K}^{\prime}
{1\over K^2+m^2}\equiv I^{\prime}(m)=I_1+2I_2m^2+...,
\nonumber\\
&
\int^{\prime}_{K_1}\int^{\prime}_{K_2}\int^{\prime}_{K_3} 
\delta (K_1+K_2+K_3)
{1\over K_1^2+m_1^2}{1\over K_2^2+m_2^2}{1\over K_3^2+m_3^2}\nonumber\\
&
\equiv H^{\prime}(m_1,m_2,m_3)=H_0+H_1{1\over 3}(m_1^2+m_2^2+m_3^2)+...
\end{eqnarray}
The prime put on the standard notations emphasizes the missing $n=0$ mode. 
The coefficients of the expansions displayed explicitly have been calculated
in \cite{jakovac} with momentum cut-off regularisation, what we also adopt 
for the present calculation. Since the propagator mass squares of the fields
are quadratic in the background field, the non-static contribution to the
static potential becomes a polynomial of $\phi^*\phi$. The ${\rm dim}>4$ terms
contribute to the potential starting from 
${\cal O}(e^6,e^4\lambda ,..,\lambda^3)$ level, 
what is negligible relative to the ${\cal O}(e^3)$ contribution of the static
screened $A_0$ field (see below). 
Therefore we truncate the non-static part at the 
quartic level. We do not write explicitly out the lengthy expression
of the regularised non-static contribution, since we concentrate on handling
non-polynomial pieces.

The static $A_0$ integration contributes the following expression to 2-loop
accuracy:
\begin{eqnarray}
&
{1\over 2}T\int_{\bf k}\ln ({\bf k}^2+M_{A_0}^2)-
{1\over 2}T\int_{\bf k}{1\over {\bf k}^2+M_{A_0}^2} [m_D^2
-e^2(2I_1+2I_2(m_H^2+m_G^2)
+...)]\nonumber\\
&
-2e^4\phi_0^2T\int_{\bf k}\int_Q^\prime{q_0^2\over Q^2(Q^2+M^2)
((K+Q)^2+m_H^2)}{1\over {\bf k}^2+M_{A_0}^2}\nonumber\\
&
-2e^2T\int_{\bf k}\int_Q^\prime{q_0^2\over (Q^2+m_H^2)((K+Q)^2+m_G^2)}
{1\over {\bf k}^2+M_{A_0}^2}.
\label{a0contr}
\end{eqnarray}
with $K\equiv (0,{\bf k}), q_0=2\pi Tn_Q, m_H^2=m^2+\lambda\Phi^2/2, m_G^2=m^2+\lambda
\Phi^2/6, M^2=e^2\Phi^2$ and $M_{A_0}^2=m_D^2+e^2\Phi^2, \Phi$ being 
the background.
The sum-integrals over the 4-momentum $Q$ in the last two integrals are hard,
therefore one can expand the corresponding propagators both in the masses 
and in the soft momentum $\bf k$. 
This technique of evaluation has been used in \cite{arzhai,zhaikast}
for pure SU(N) gauge theory. The leading
contribution is arrived at by replacing in (\ref{a0contr}) $K$ by 0.
Adding just these contributions from the last two integrals to the first two terms
one finds the non-polynomial part of the potential energy:
\begin{equation}
-{T\over 12\pi}(m_D^2+e^2\Phi^2)^{3/2}+e^2({\Lambda T\over 2\pi^2}-{TM_{A_0}
\over 4\pi})(-{\Lambda T\over 2\pi^2}+{T^2\over 6}-{1\over 2e^2}m_D^2).
\label{nonpol1}
\end{equation}
In addition also divergent pieces $\sim\Phi^2$ appear.
The further contributions in the last two integrals of (\ref{a0contr})
do not contribute cut-off independent finite terms to the potential 
energy to ${\cal O}(e^4,e^2\lambda ,\lambda^2)$. The polynomial divergent 
contributions will not be displayed here, since
we concentrate on the consistency of the treatment of non-polynomial
terms in the Lagrangian.

When in (\ref{nonpol1}) one uses the fact that $m_D^2=e^2T^2/3$,
 the effect of the $A_0$-integration can be summarised as 
\begin{equation}
U_{nonpol}(\phi )=-{e^3T\over 12\pi}({1\over 3}T^2+\phi^*\phi )^{3/2}+
{e^3\Lambda T^2\over 8\pi^3}({1\over 3}T^2+\phi^*\phi )^{1/2}.
\end{equation}

A more compact form of the renormalised potential emerges if the 
renormalisation conditions to be applied to the complete expression refer
to the T-independent part of the finite T expression rather than
directly to the T=0 expression of the potential:
\begin{equation}
{\partial^2U(\phi ,{\rm T-indep})
\over \partial\phi^2}_{|\phi =0}=m^2,~~~{\partial^4U(\phi ,{\rm T-indep})
\over \partial\phi^4}_{|\phi =0}=\lambda.
\label{renorm}
\end{equation}
This normalisation scheme simplifies the comparison of different approximations
to the effective theory. On the other hand the relation of the 
renomalised mass parameter to some physical scale at $T=0$ becomes 
 more complicated.
For instance the expression of the expectation value 
of the scalar field at $T=0$ 
calculated with 1-loop accuracy using counterterms derived from (\ref{renorm})
reads as
\begin{equation}
v^2=-{6m^2\over \lambda}[1+{\lambda\over 8\pi^2}\{(1+\log C)(1+18
{e^4\over\lambda^2})+{1\over 2}\log (-{\mu^2\over 2m^2})+{9e^4\over \lambda^2}
\log (-{\mu^2\lambda\over 6e^2m^2})\}].
\label{vev}
\end{equation}
The scale $\mu$ is the renormalisation scale, which below will be chosen
to be T, $C=2\pi\exp (-\gamma_E)$.

The final result is (without giving explicitly the functions $h_i
(e^2,\lambda,..)$ below):
\begin{eqnarray}
&
L_{3D}^{pot}={1\over 2}m_T^2\varphi^*\varphi +{1\over 24}\lambda_3
(\varphi^*\varphi )^2-{e_3^2\over 12\pi}Q^3(\varphi )+
{e_3^3\Lambda\over 8\pi^3}Q(\varphi )\nonumber\\
&
+(h_1\Lambda+h_2T\log ({\Lambda\over T})+h_3\Lambda\log ({\Lambda\over T}))
\varphi^*\varphi
\label{lpot}
\end{eqnarray}
with $\varphi =\phi /\sqrt{T}$ and
\begin{eqnarray}
&
m_T^2=m^2+T^2\{{1\over 12}(3e^2+{2\over 3}\lambda )+
{1\over \pi^2}[e^4({1\over 16}+{1\over 96}\log C+k_1)\nonumber\\
&
+\lambda^2({5\over 864}+
{7\over 864}\log C+k_2)+k_3e^2\lambda ]-{1\over 16\pi^2}\log {\mu\over T}
(e^4+{5\lambda^2\over 54})\},\nonumber\\
&
\lambda_3=\lambda T,~~~Q(\varphi )=({1\over 3}T+\varphi^*\varphi )^{1/2},
~~~e_3^2=e^2T.
\end{eqnarray}
The regularisation dependent constants $k_1,k_2,k_3$ have the values:
\begin{equation}
k_1=0.473515,~~~k_2=0.0190808,~~~k_3=-0.0901793.
\end{equation}

The kinetic terms of the static $a_i(x)=A_i(x)/\sqrt{T}$ and $\varphi (x)$
fields can be found by studying the contribution of the integrated out 
fields to the corresponding 2-point functions. As it has been shown by 
\cite{fodor1,fodor2}, for the calculation of the 2-loop effective action one 
needs only the ${\cal O}(e)$ T-dependent 1-loop corrections of the wave function
rescaling factor. In accordance with their conclusion our explicit 
calculation shows that the contribution from non-static modes is 
${\cal O}(e^2)$ (including the piece necessary in the 4D renormalisation).
Therefore only the static $A_0$ integration should be taken into account in 
$\delta Z_{\phi^2}^T$. If one terminates the gradient expansion with the 
usual kinetic term one finds:
\begin{eqnarray}
&
L_{3D}^{kin}={1\over 4}f_{ij}f_{ij}+{1\over 2}(d_i\varphi )^*(d_i\varphi )
+{1\over 2}\delta Z_{\phi_H}^{T}[(\partial_i\varphi_H)^2+e^2a_i^2\varphi_H^2],
\nonumber\\
&
\delta Z_{\phi_H}^T=-{e^4\Phi^2T\over 48\pi M_{A_0}^3}\sim {\cal O}(e).
\label{lkin1}
\end{eqnarray}
This equation shows that only the Higgs part of the complex scalar field 
receives T-dependent rescaling. The lower case symbols refer to 3D 
quantities.

Another alternative, which we are going to claim to be the correct procedure,
is to retain the full non-local $A_0$ contribution to the Higgs 2-point
function:
\begin{eqnarray}
&
L_{3D}^{kin}={1\over 4}f_{ij}f_{ij}+{1\over 2}(d_i\varphi )^*(d_i\varphi )+
{1\over 2}\int_{\bf k}\varphi_H ({\bf k})\varphi_H (-{\bf k})[{\cal M}({\bf k})-
{\cal M}(0)],\nonumber\\
&
{\cal M}({\bf k})=-2e^4\Phi^2T\int_{\bf p}\int_{\bf q}
{1\over p^2+M_{A_0}^2}{1\over q^2+M_{A_0}^2}\delta ({\bf k+p+q}).
\label{lkin2}
\end{eqnarray}
(${\cal M}(0)$ is subtracted since the corresponding mass-contribution is
already contained in (\ref{lpot})). Since ${\cal M}$ is already ${\cal O}(e^4)$
the terms completing  the non-local piece to be 
gauge invariant can be neglected.

The sum of (\ref{lpot}) and of either (\ref{lkin1}) or (\ref{lkin2}) represents
two alternatives of the 3D reduced model derived from the full theory
with 2-loop accuracy.
\vskip 1truecm
3. For the 2-loop computation of the effective potential in the reduced model
one needs the propagator, and the cubic and quartic vertex parts from
the reduced lagrangian, calculated on a constant background $\varphi_0$.
The formulae will be presented for the non-local case (\ref{lkin2}).
The local approximation will be commented in the discussion part.
\begin{eqnarray}
&
L_{3D}={1\over 4}f_{ij}f_{ij}+{1\over 2}((\partial_i\varphi_H)^2+(\partial_i
\varphi_G)^2+m_H^2\varphi_H^2+m_G^2\varphi_G^2)+{1\over 2}e_3^2
\varphi_0^2a_i^2\nonumber\\
&
+e_3^2\varphi_0\varphi_Ha_i^2+ie_3a_i(\varphi_H\partial_i\varphi_G-
\varphi_G\partial_i\varphi_H)+q_{111}\varphi_H^3+q_{122}\varphi_G^3
\nonumber\\
&
+{1\over 2}e_3^2a_i^2(\varphi_H^2+\varphi_G^2)+{1\over 24}(\lambda_{11}
\varphi_H^4+2\lambda_{12}\varphi_H^2\varphi_G^2+\lambda_{22}\varphi_G^4)
\nonumber\\
&
+{1\over 2}\int_{{\bf k}}\varphi_H({\bf k})
\varphi_H(-{\bf k}){\cal M}({\bf k})+L_{3D,ct}
\end{eqnarray}
with
\begin{eqnarray}
&
m_G^2=m_T^2+{\lambda_3\over 6}\varphi_0^2-{e_3^3\over 4\pi}Q,~~~m_H^2=m_G^2+
{\lambda_3\over 3}\varphi_0^2,\nonumber\\
&
q_{122}={\lambda_3\over 6}\varphi_0-{e_3^3\varphi_0\over 8\pi}Q^{-1},
~~~q_{111}=q_{122}-{e_3^3\varphi_0^3\over 24\pi}Q^{-3}\nonumber\\
&
\lambda_{22}=\lambda_3-{3e_3^3\over 4\pi}Q^{-1},~~~\lambda_{12}=\lambda_{22}
+{3e_3^3\over 4\pi}\varphi_0^2Q^{-3},\nonumber\\
&
\lambda_{11}=\lambda_{12}+
{3e_3^3\varphi_0^2\over 4\pi}(Q^{-3}-\varphi_0^2Q^{-5}).
\label{param}
\end{eqnarray}
The ${\cal O}(e^3)$ expression of the effective potential in 3D Landau gauge
is the following:
\begin{eqnarray}
&
U_{3D}^{finite}={1\over 2}m_T^2\varphi_0^2+{\lambda_3\over 24}\varphi_0^4
-{e_3^3\over 12\pi}Q^{3}-{1\over 6\pi}[(e_3^2\varphi_0^2)^{3/2}+{1\over 2}
(m_G^3+m_H^3)],\nonumber\\
&
U_{3D}^{div}={\Lambda\over 2\pi^2}[e_3^2\varphi_0^2+{1\over 2}(m_G^2+m_H^2)]
+{\rm tree-level~ "counterterms"}
\label{loc1loop}
\end{eqnarray}
When the above expressions for $m_G^2,m_H^2$ are substituted into the divergent 
part one easily checks the exact cancellation of the polynomial 
($\sim\varphi_0^2$) and of the non-polynomial $(\sim M_{A_0}T)$ divergencies.
This makes explicit the consistency of the treatment of the reduced theory
with non-polynomial potential to ${\cal O}(e^3)$ accuracy.

The local 2-loop contributions come from the same set of Feynman diagrams
like the one used for the PI-step. This time the two standard integrals
are three-dimensional, and were calculated
with cut-off regularisation already in \cite{jakovac}:
\begin{eqnarray}
&
I_3(m)={\Lambda\over 2\pi^2}-{m\over 4\pi}\nonumber\\
&
H_3(m_1,m_2,m_3)={1\over 16\pi^2}(\log {\Lambda\over\mu}+
\log {3\mu\over m_1+m_2+m_3}+L_0),
\end{eqnarray}
$(L_0=1.0585301-\log 3 )$. 
When choosing (as finally we did in the PI-step too) $\mu =T$, 
the finite part of the 2-loop local effective potential 
contribution has the following form:
\begin{eqnarray}
&
U_{2-loop}^{loc}=\nonumber\\
&
{e_3^2\over 32\pi^2}[{1\over 2}m_H^2+2Mm_H+Mm_G+m_Hm_G-M^2-{m_H-m_G\over M}
(m_H^2-m_G^2)]\nonumber\\
 &
+{1\over 384\pi^2}[3\lambda_{22}m_G^2+2\lambda_{12}m_Hm_G+3
\lambda_{11}m_H^2]\nonumber\\
&
{e_3^2\over 16\pi^2}[{m_H^4\over 4M^2}\log(1-({M\over M+m_H})^2)+
{(m_H^2-m_G^2)^2 \over 2m^2}\log(1+{M\over m_H+m_G})\nonumber\\
&
+m_H^2\log {M+m_H\over 2M+m_H}-{M^2\over 2}
\log{(M+m_H)(M+m_H+m_G)^3\over (2M+m_H)^4}]\nonumber\\
&
{e_3^2\over 16\pi^2}(m_H^2+m_G^2-2M^2)(L_0-\log{M+m_G+m_H\over 3T})
+{q_{122}^2\over 16\pi^2}(L_0-\log{2m_G+m_H\over 3T})\nonumber\\
&
-{3q_{111}\over 16\pi^2}(L_0-\log{2m_H\over 3T})
\label{loc2loop}
\end{eqnarray}
($M\equiv e_3\varphi_0$).
One adds  to this expression the non-local contribution:
\begin{equation}
U_{2-loop}^{non-loc}=-{e_3^4\varphi_0^2\over 16\pi^2}(L_0-\log{2M_{A_0}+m_H
\over 3T}).
\label{nloc2loop}
\end{equation}
If the contribution of the static $A_0$ field would be approximated locally
(\ref{lkin1}), the Higgs-mass $m_H^2$
 would receive an extra contribution, destroying
the ${\cal O}(e^3)$ cancellation of the non-polynomial  divergences. The
"counterterm" proportional to $\delta Z_{\phi}^T$ would also contribute at 
1-loop, corresponding to the expansion of (\ref{nloc2loop}) 
in powers of $m_H/M_{A_0}$. These remarks strongly favor the use of the 
complete non-local reduced model when arbitrary high spatial momenta 
$(<\Lambda)$ are allowed.

The exact form of $I_3(m)$ does not contain any $\sim\Lambda^{-1}$ part,
therefore the cut-off dependent part of the reduced Lagrangian cannot
produce at 1-loop any finite piece. Therefore the finite part of the 
sum of (\ref{loc1loop}),(\ref{loc2loop}) and (\ref{nloc2loop}) gives
our final result for the 2-loop finite temperature effective potential 
of the scalar electrodynamics. When compared with \cite{hebecker1,hebecker2}
one finds complete formal 
agreement of the terms independent of the regularisation
and of the normalisation conditions. But one should not forget that
the couplings and masses appearing in it are given by (\ref{param}),
what is different from the solution of the self-consistent Schwinger-Dyson
equations. Still the structural agreement represents a
 satisfactory evidence for the 
correctness of our reduced partially integrated model and its perturbative
treatment described above.

In order to investigate quantitatively the importance or negligibility
of the non-polynomial contributions we evaluate the above
expression also in a polynomial (superrenormalisable) approximation.
It corresponds to expanding in the potential energy of the reduced model the
term $\sim M_{A_0}^3$ in powers of $\varphi^*\varphi$ up to the quartic piece 
and dropping all the higher dimensional operators \cite{farakos1,jako1}. A
small new feature is that now also the cut-off dependent non-polynomial
term in (\ref{nonpol1}) is expanded, what modifies the induced counterterm
structure.
This leads to the following set for the finite parts of the couplings:
\begin{eqnarray}
&
m_G^2=m_T^2+{\lambda_3\over 6}\varphi_0^2-{e_3^3\tilde m_D\over 4\pi},
~~~m_H^2=m_G^2+{\lambda_3\over 3}\varphi_0^2,\nonumber\\
&
q_{111}=q_{12}={\lambda_3\over 6}\varphi_0-{e_3^3\varphi_0\over 8
\pi\tilde m_D},~~~\lambda_{11}=\lambda_{12}=\lambda_{22}=\lambda_3-{3e_3^3\over
4\pi\tilde m_D},
\end{eqnarray}
($\tilde m_D\equiv (T/3)^{1/2}$).

\begin{figure}[t]
\epsfig{file=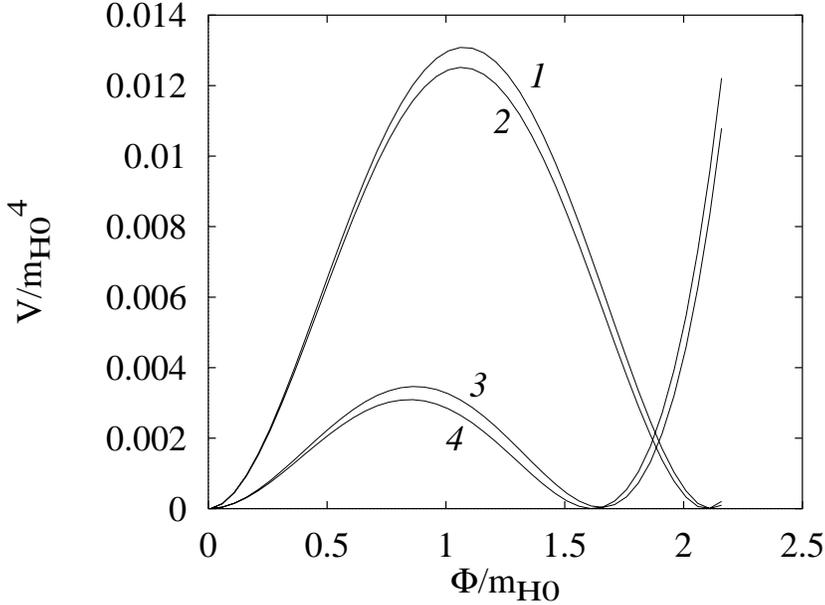,height=8truecm}
\caption{\em The effective potential at the respective critical temperatures
1. at one loop in polynomial approximation
2. at one loop in nonpolynomial approximation
3. at two loop in polynomial approximation
4. at two loop in nonpolynomial approximation}
\label{fig:1}
\end{figure}

In Fig.~\ref{fig:1} we present the effective potential in four different
approximations at
the respective transition temperatures ($e=2/3,~\lambda =0.3)$. 
The potential and the field variable are both scaled by appropriate powers of
$m_{H0}^2=-2m^2$.
The four curves represent the
potential from 1-loop and 2-loop approximations computed with 
superrenormalisable and with non-polynomial potentials each.
The curves show the same qualitative feature as found by 
\cite{hebecker1,hebecker2}. The application of our renormalisation 
conditions does not change the conclusion that perturbation theory
is not reliable for such large value of the coupling $\lambda$. 
In this specific point $v\approx 1.18(-6m^2/\lambda )^{1/2}$, what 
leads to quantitative agreement with the results of \cite{hebecker1,hebecker2}.
The 1-loop correction to $v_0$ is fully dominated 
by the first term of the curly bracket
of (\ref{vev}) $\sim e^4/\lambda^2$. 

The effect of 
the higher dimensional operators is quite noticeable both on the $e^{3/2}$
and the $e^4$ level. They tend to smoothen the phase transition. Still
the result of the simpler superrenormalisable version in this case seems 
to be quite satisfactory. One might like to
 characterize the effective potential at
the transition with a single number: the interface tension, calculated
in thin wall approximation. It is notable that the absolute value of the 
difference between the polynomial and non-polynomial versions is the same
both in 1-loop and 2-loop approximations. When measured in units of
$(-2m^2)^{3/2}$ one finds:
\begin{eqnarray}
&
\sigma (pol,1-loop)=.227803,~~\sigma (non-pol,1-loop)=.220705\nonumber\\
&
\sigma (pol,2-loop)=.090203,~~\sigma (non-pol,2-loop)=.083323
\end{eqnarray}
The percentual importance of the
higher dimensional operators grows to 8\% at two loop, since the surface tension
decreases to more than the half of its 1-loop value.

In conclusion it is clear
that such analysis will improve our understanding of the physics 
of the electroweak phase transition too.
\vskip .6truecm
{\bf Acknowledgements}
We are grateful for enjoyable discussions with Z. Fodor, I. Montvay and M.
Shaposhnikov in the stimulating atmosphere of the 
CERN Theory Division. Also the grant of the Hung. Science 
Foundation is gratefully acknowledged.

\end{document}